\documentclass[prl,twocolumn,showpacs,superscriptaddress,nofootinbib]{revtex4-2}
\pdfoutput=1
\usepackage{graphicx}
\usepackage{epstopdf}
\usepackage{bm}
\usepackage{amssymb}
\usepackage{color}
\usepackage{amsmath}
\usepackage{amstext}
\usepackage{latexsym}
\usepackage[usenames,dvipsnames]{xcolor}
\usepackage[colorlinks=true,citecolor=MidnightBlue,linkcolor=RubineRed,urlcolor=MidnightBlue]{hyperref}
\usepackage{float}

\def\x{{\rm\bf x}}

\newcommand{\beq}{\begin{equation}}
\newcommand{\eeq}{\end{equation}}
\newcommand{\beqa}{\begin{eqnarray}}
\newcommand{\eeqa}{\end{eqnarray}}

\usepackage{tikz,xcolor,hyperref}
\definecolor{lime}{HTML}{A6CE39}
\DeclareRobustCommand{\orcidicon}{
\begin{tikzpicture}
\draw[lime, fill=lime] (0,0)
circle[radius=0.16]
node[white]{{\fontfamily{qag}\selectfont \tiny \.{I}D}}; 
\end{tikzpicture}
\hspace{-2mm}
}
\foreach \x in {A, ..., Z}{%
\expandafter\xdef\csname orcid\x\endcsname{\noexpand\href{https://orcid.org/\csname orcidauthor\x\endcsname}{\noexpand\orcidicon}}
}

\begin{document}

\title{Non-equilibrium Dynamics and Universality of 4D Quantum Vortices and Turbulence}

\author{Wei-Can Yang\hspace{-1.5mm}\orcidA{}
}\email{weicanyang@outlook.com}
\affiliation{Yukawa Institute for Theoretical Physics, Kyoto University, Kyoto 606-8502, Japan}

\begin{abstract}
The study of quantum vortices provides critical insights into non-equilibrium dynamics across diverse physical systems. While previous research has focused on point-like vortices in two dimensions and line-like vortices in three dimensions, quantum vortices in four spatial dimensions are expected to take the form of extended vortex surfaces, thereby fundamentally enriching dynamics. Here, we conduct a comprehensive numerical study of 4D quantum vortices and turbulence. Using a special visualization method, we discovered the decay of topological numbers that does not exist in low dimensions, as well as the high-dimensional counterpart of the vortex reconnection process. We further explore quench dynamics across phase transitions in four dimensions and verify the applicability of the higher-dimensional Kibble-Zurek mechanism. Our simulations provide numerical evidence of 4D quantum turbulence, characterized by universal power-law behavior.  These findings reveal universal principles governing topological defects in higher dimensions, offering insights for future experimental realizations using synthetic dimensions.
\end{abstract}

\maketitle

Quantum vortices are among the most striking manifestations of macroscopic quantum phenomena. As quantized topological defects, they govern the complex dynamics of a wide range of non-equilibrium quantum systems. Initially discovered in superfluid helium and later observed in ultracold atomic gases, quantum vortices have emerged as a unifying concept across many-body physics. They not only form the structural backbone of quantum turbulence in superfluids \cite{Salomaa_1987,Nemirovskii_1995,Vinen_1968,Yang_2023Motion} and flux lines in type-II superconductors \cite{Wallraff_2003,Fazio_2001,Abrikosov_2004}, but also appear in a wide variety of physical systems ranging from nonlinear optics \cite{Drori_2023} and magnetically confined plasmas \cite{Shukla_2006,Mahajan_2011} to early-universe cosmology \cite{Kibble_1997,Kleinert_1999}.
 
The investigation of quantum vortices has spanned decades, yielding a wealth of universal phenomena and laws, particularly in the study of quantum turbulence formed by a multitude of vortices \cite{Vinen_2002,Barenghi_2014,Yang_2024Spontaneous}. Quantum turbulence represents the quantum counterpart to classical turbulence --- the last unresolved puzzle in classical physics \cite{Sreenivasan_1999}. Research into quantum turbulence has unveiled key microscopic mechanisms underlying turbulent behavior, with the primary processes being the excitation of Kelvin waves and vortex reconnection \cite{Minowa_2022,Yang_2025quantumturbulence,Minowa_2025}. In three-dimensional(3D) systems, these mechanisms facilitate the transfer of energy from large scales to small scales, ultimately dissipating through phonon excitations \cite{Chen_2003,Muller_2020}. Within the inertial range of this energy cascade, the energy spectrum aligns perfectly with the Kolmogorov prediction of a $k^{-5/3}$ power-law \cite{Kobayashi_2005,Kobayashi_2006,Yepez_2009,Yang_2025quantumturbulence,Zeng_2025}, thereby establishing a remarkable bridge between quantum and classical complex physics.

Quantum vortices exhibit dimension-dependent structures and dynamics, profoundly influencing the universality of quantum turbulence. As mentioned above, in 3D systems, quantum vortices appear as line-like topological defects, facilitating a forward energy cascade through vortex reconnections and Kelvin wave excitations. In contrast, in 2D systems, quantum vortices are point-like defects whose intricate interactions lead to the formation of vortex lattices \cite{Abo_Shaeer_2001,Fischer_2003,Kasamatsu_2003,Yang_2020,Yang_2021} or Onsager vortex clusters, characterized by negative-temperature states and an inverse energy cascade, transferring energy from small-scale to larger-scale coherent structures \cite{Onsager_1949,Simula2014,Johnstone_2019,Gauthier_2019,Valani_2018,Groszek2016,Yang_2024Cluster,Yang_2025quantumturbulence}. Such significant differences between 2D and 3D quantum vortices, both in topology and energy transfer direction, naturally motivate the exploration of vortex dynamics and turbulence in higher-dimensional systems. Recent studies have begun to investigate simple vortex structure and interaction in 4D spaces, revealing richer degrees of freedom and more complex phenomena \cite{McCanna_2021,McCanna_2024,Middletonspencer2024}.

\begin{figure*}[t]
\centering
\includegraphics[width=19cm,trim=0 80 0 0]{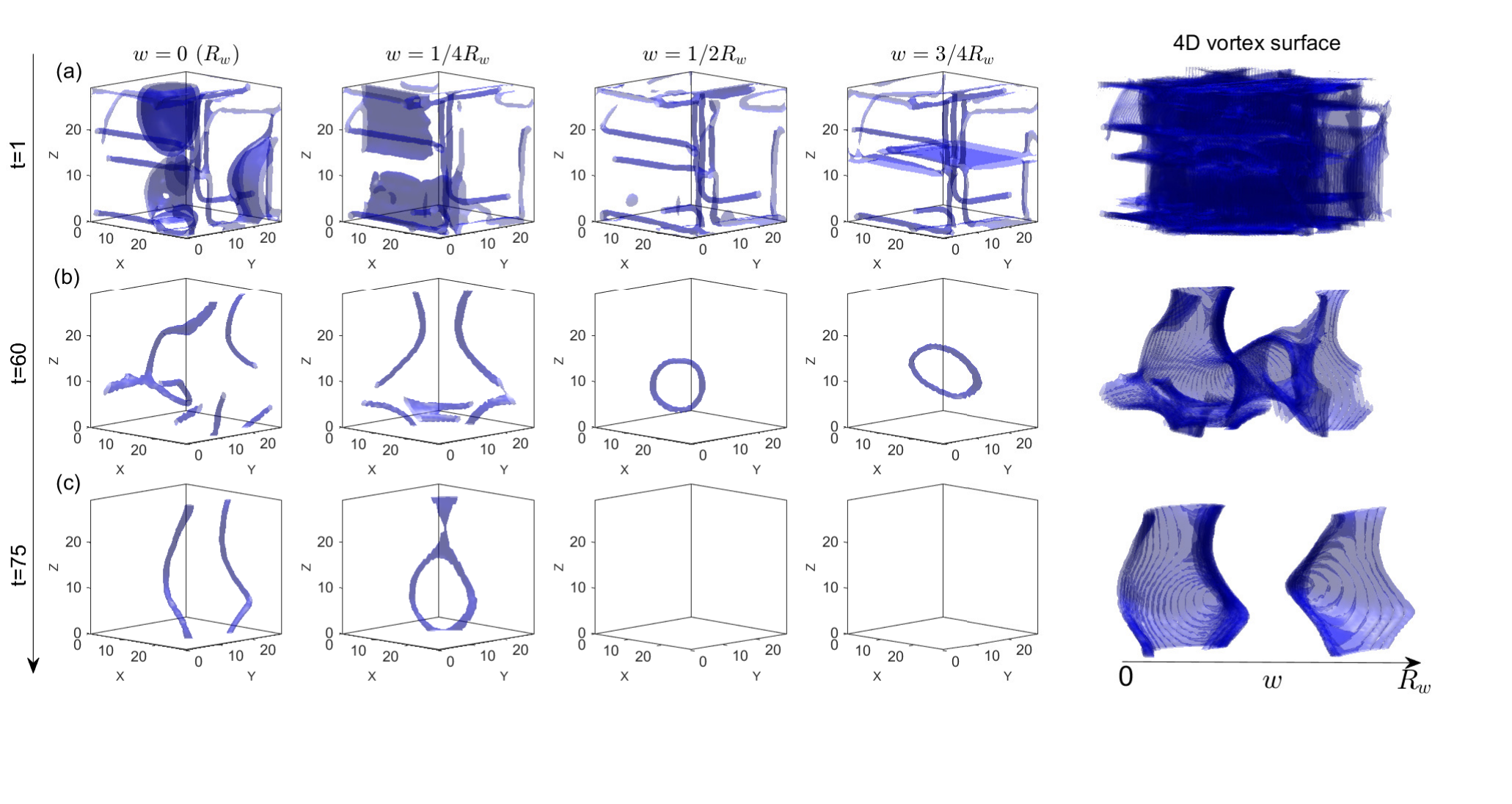}
\caption{The evolution of 4D vortex dynamics. From top to bottom, each row shows snapshots at three characteristic times $t=1$, $t=60$ and $t=75$. In each row, the first four panels display three-dimensional cross-sections at positions \( w = 0 \ (R_w),\ R_w/4,\ R_w/2,\ 3R_w/4 \) along the additional fourth spatial dimension \( w \). The rightmost panel shows the reconstructed 4D vortex surface by stacking all cross-sections along the '$w$ -direction'. In the second row (\( t = 60 \)), three topological holes are clearly visible, corresponding to a genus of 3. In the third row (\( t = 75 \)), the structure reduces to a cylindrical shape with genus 1.
A more detailed dynamic process can be found in the supplementary video \cite{Movie}.}
\label{Evolution}
\end{figure*}

The pronounced differences between 2D and 3D quantum turbulence raise a fundamental question: How do the structures and dynamics of quantum vortices evolve when extended to higher dimensions? Can such systems still support turbulence state? And if so, do they obey universal scaling laws akin to those found in lower dimensions?

Motivated by recent advancements in synthetic dimension --- extra degrees of freedom engineered by coupling internal states of particles \cite{Boada_2012,Price_2017}, exploring 4D quantum vortices and turbulence becomes not only theoretically intriguing but also experimentally possible \cite{Bouhiron_2024,Ozawa_2019}. 
Recent experiments have successfully simulated 4D critical phenomena using synthetic temporal modulations \cite{Madani_2025}. Furthermore, investigations into higher-dimensional vortices may provide insights into cosmological scenarios, as theories of the early universe, including brane-world models and higher-dimensional cosmologies, suggest that extended topological defects and turbulence could naturally occur during early cosmological phase transitions \cite{Anabalon_2008,Prokopec_1991,Copeland_1989}.

In this Letter, we address these fundamental questions by systematically investigating the dynamical behaviors of quantum vortices and turbulence in four spatial dimensions through numerical simulations. We reveal, for the first time, clear signatures of 4D quantum turbulence, identify universal scaling laws, including vortex decay, energy spectra, and distinct velocity distribution tails. We also verify the higher-dimensional applicability of the Kibble-Zurek mechanism, demonstrating the robustness of extended two-dimensional topological defects in four dimensions.

Higher-dimensional vortex dynamics can be effectively explored by generalizing the mean-field approach commonly used for Bosonic quantum fluids \cite{McCanna_2021,McCanna_2024,Middletonspencer2024}. In this work, we utilize the widely adopted and extensively validated stochastic Gross–Pitaevskii equation(SGPE) \cite{Cockburn_2009,Thudiyangal_2024,Savenko_2013}, extending it to four spatial dimensions, read as
\begin{equation}
(i - \gamma)\frac{\partial \phi({\bf r},t)}{\partial t} 
= (-\frac{1}{2}\nabla^2 -\mu + g|\phi({\bf r},t)|^2)\phi({\bf r},t)+ \eta({\bf r},t)
\end{equation}
where ${\bf r}=(x,y,z,w)$ is the four-dimensional coordinate, $\phi$ is the condensed wave function, $\gamma$ the dissipation rate, $\mu$ the chemical potential and $g$ the interacting strength.
 The thermal noise given by the fluctuation-dissipation theorem $\langle \eta({\bf r}, t) \eta^*({\bf r}', t') \rangle = 2 \gamma T \delta({\bf r} - {\bf r}') \delta(t - t')$, where $T$ is the temperature. 
We always fix the parameter $\gamma=0.1$, $g=1$ and $T=10^{-6}$. The chemical potential $\mu$ is adjustable, enable the system to be linear quenched, which we will employ later.

\begin{figure}[t]
\centering
\includegraphics[width=9cm,trim=20 0 0 0]{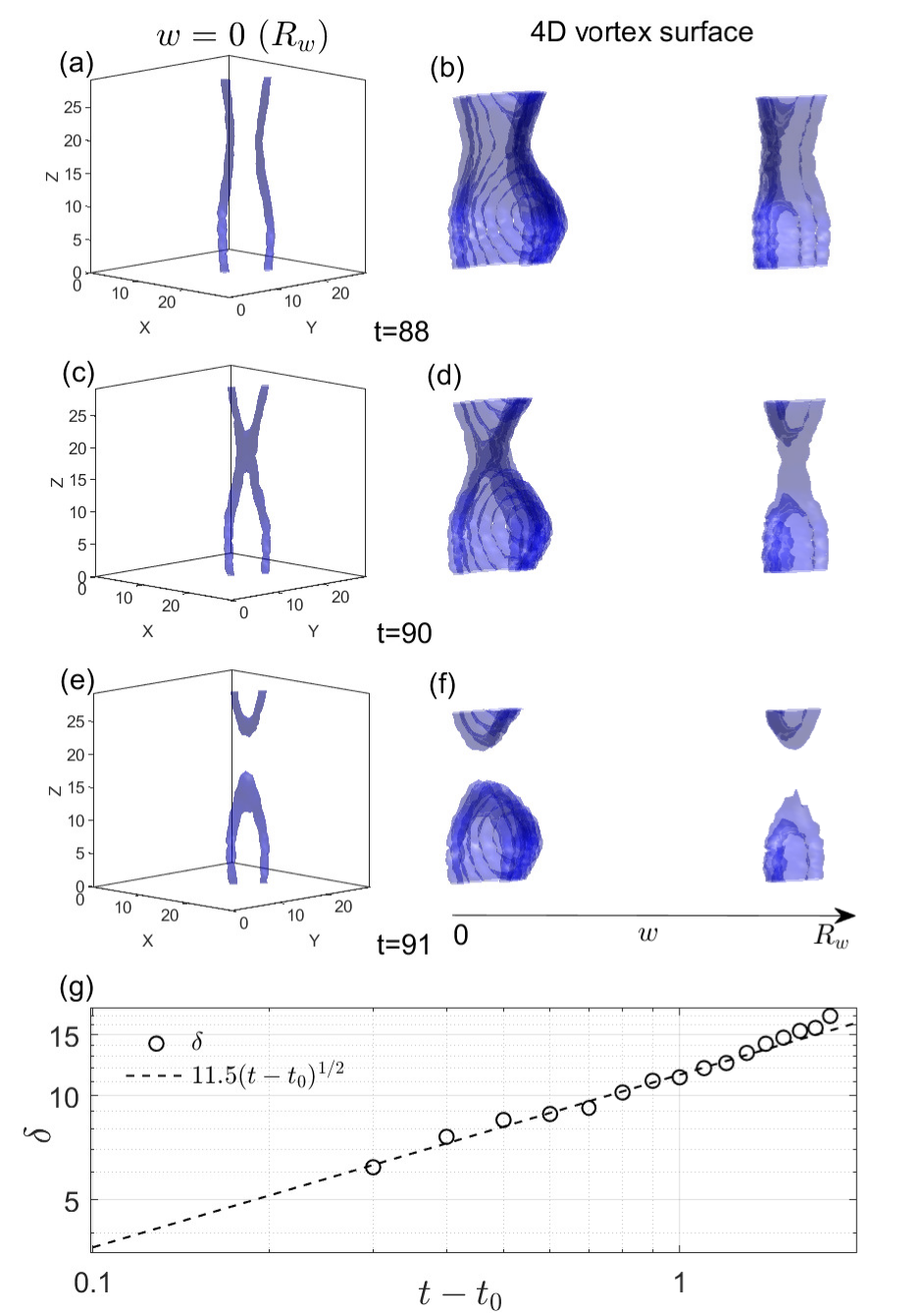}
\caption{Dynamics of the final vortex surface splitting. Each row shows snapshots at different times. The left column presents 3D cross-sections at \(w = 0\ (R_w)\), while the right column displays the reconstructed 4D vortex surfaces. The reconnection of vortex lines in 3D corresponds exactly to the splitting of the vortex surface in 4D. In 3D, two vortex lines reconnect to form a vortex ring; in 4D, a cylindrical vortex surface splits into a spherical surfaces, reducing the topological number (genus) from 1 to 0. 
Panel (g) shows the time evolution of the distance \(\delta\) between the closest vortex points during the split process. The result follows the characteristic square-root scaling law of vortex reconnection: \(\delta(t) \propto (t - t_0)^{1/2}\).
}
\label{reconnection}
\end{figure}

In our simulations, we employ a four-dimensional square periodic boundary conditions with size of $L=30$.
To investigate the structure and interaction of vortices, we first fix the chemical potential at 
$\mu=10$ and initially consider a wave function with uniform density $\phi({\bf r},t=0)=\phi_i$. We know that a four-dimensional system has six independent directions, in each of these directions we employ the phase imprint method to randomly place two vortex planes, as illustrated in Fig.\ref{Evolution} (a). 

In the additional dimension $w$, we present four uniformly spaced "cross-sections", which appear as three-dimensional systems by the remaining dimensions $x$, $y$, and $z$. Within these cross-sections, vortices exhibit two distinct states: surface-like and line-like. In fact, the line-like state emerges as a result of the vortex surface being sliced along the  $w$-dimension. 
By extending and superimposing each cross-section in the $w$-direction along a hypothetical direction $(x,y)/\sqrt{2}$, we can obtain a vortex surface distribution that closely approximates the actual result. 
In reality, the orthogonality of the fourth dimension $w$ to the other three dimensions ($x$, $y$, and $z$) makes it impossible to fully represent the system in a strictly accurate manner. Despite this limitation, the vortex surfaces obtained through our method accurately reflect the true distribution in higher dimensions, effectively visualizing the dynamics and providing clear insights into how vortex surfaces move and evolve in four-dimensional space \cite{Supplementary_material}.

As illustrated in the second and third rows of subfigures (b,c) in Fig.\ref{Evolution}, vortices gradually decay over time due to excitation and dissipation. After the number of vortices on each cross-section decreases, the overall structure of the vortex surface becomes clearly discernible. Surprisingly, all vortex surfaces merge into a single connected surface, distinct from isolated vortices typically observed in 2D or 3D systems. The decay process of the entire vortex surface resembles the shrinking behavior observed in dissipating vortex rings. Although vortex surface splitting events do occur, they do not disrupt the global connectivity of the vortex surface.

We propose that the decay of the vortex surface correlates directly with the reduction of its topological number, namely its genus($g$) \cite{What_2025}. For example, at time $t=60$, we have relatively complex structure with $g=3$ (Fig.\ref{Evolution}(b)); at $t=75$, it decreases to a cylindrical shape with $g=1$ (Fig.\ref{Evolution}(c)); and by $t=91$, the surface further decays to a spherical topology with $g=0$ (Fig.\ref{reconnection}(f)), eventually disappearing via a shrinking process. Each splitting event of the vortex surface results in a reduction of genus by one, thereby altering the overall topological structure.

In Fig.\ref{reconnection}, we explicitly demonstrate the final vortex surface splitting event, capturing the transition from a cylindrical surface (genus=1) to a spherical surface (genus=0). On the cross-section at $w=0$ (or $R_w$ due to the periodic boundary), this transition appears as vortex line reconnection. The reconnection process closely follow the equation of intervortex distance \cite{Minowa_2022}: 
\begin{equation}
    \delta(t)=A\sqrt{\kappa\ |t-t_0|\ }
\end{equation}
where $\delta$ is the intervortex distance, $A$ is a dimensionless amplitude factor and $\kappa$ is the circulation quanta $h/m$.

In 3D quantum turbulence, vortex reconnections play a crucial role by altering vortex line topology and transferring energy either to the normal fluid through mutual friction or by emitting sound waves and Kelvin waves. In our 4D scenario, vortex surface splitting corresponds directly to these reconnection events, resulting in an irreversible reduction of the topological number (genus) and a similar mechanism of energy transfer.

Having comprehensively described the dynamic processes and decay behavior of 4D vortex surfaces above, we are motivated to investigate whether universal scaling behaviors observed in lower-dimensional quantum turbulence persist in the richer scenario of 4D systems.
To this end, we now turn to exploring quench dynamics—an effective approach to generating and studying turbulence states by driving the system through a phase transition \cite{Saito_2007,Del_2014,Chesler_2015,Beugnon_2017,Uhlmann_2007,Uhlmann_2010,Uhlmann_2010_2,Supplementary_material}. Specifically, we consider a quench process induced by a linear variation of the chemical potential across the critical point $\mu_c=0$ with equation
\begin{equation}
\mu(t) = (\mu_f - \mu_0)\frac{t}{\tau_Q} + \mu_0 \quad \text{for } 0 \le t \le \tau_Q,
\end{equation}
where $\mu(t)$ increases linearly from $\mu_0=-0.1$ to the maximum value $\mu_f$ at $t = \tau_Q$. 

According to the Kibble-Zurek mechanism, when the system is driven across a phase transition, the diverging relaxation time near the critical point leads to a freeze-out of the dynamics. The system ceases to follow the instantaneous ground state and reenters quasi-adiabatic evolution after a characteristic freeze-out time $\hat{t}$ determined by the quench rate. This freeze-out time sets the effective correlation length at the transition, which in turn determines the density of topological defects formed during the quench \cite{Kibble_1976,Zurek_1985,Kibble_1997,Donadello_2016}. As a result, the defect density 
$n$ scales with the quench time $\tau_Q$ as
\begin{equation} 
n \sim \frac{\hat{\xi}^{d}}{\hat{\xi}^{D}} \sim \tau_Q^{-\frac{(D-d)\nu}{1+\nu z}}, 
\end{equation}
where $\hat{\xi}$ is the freeze-out correlation length, 
$d$ is the topology defect dimension, $D$ is the spatial dimension, $\nu$ is the critical exponent for the correlation length, and 
$z$ is the dynamic critical exponent.

\begin{figure}[t]
\includegraphics[width=9.5cm,trim=20 20 0 10]{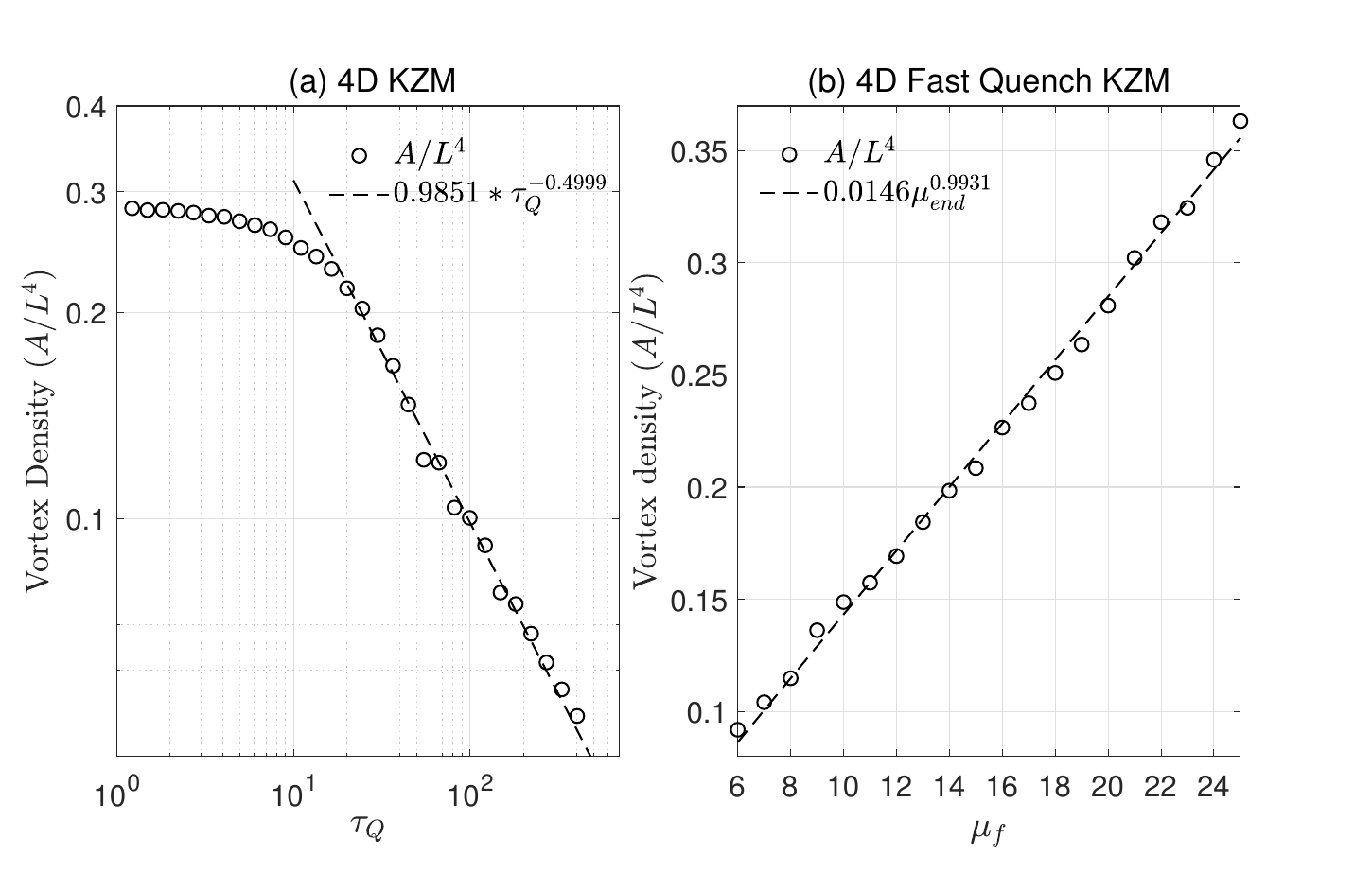}
\caption{(a) Vortex surface density as a function of quench time \(\tau_Q\) with fixed $\mu_f=20$, showing a power-law scaling consistent with the Kibble-Zurek mechanism (KZM) prediction of exponent \(-1/2\).  
(b) Vortex surface density after instantaneous quench as a function of the final chemical potential \(\mu_f\), exhibiting a scaling behavior consistent with the fast-quench KZM prediction of exponent \(+1\).}
\label{4DKZM}
\end{figure}

In the 4D bosonic system undergoing a second-order phase transition in the mean-field regime, we have $D=4$, $\nu=1/2$ and $z=2$, then the final power law depends on the dimension of the topological defect with $n\sim\tau_Q^{(d-4)/4}$. If the system supports only vortex surfaces as topological defects, then we have $d=2$ and the density of vortex surfaces is expected to follow a universal power-law decay with exponent $-1/2$. 

\begin{figure}[t]
   \includegraphics[width=9cm,trim=0 30 0 0]{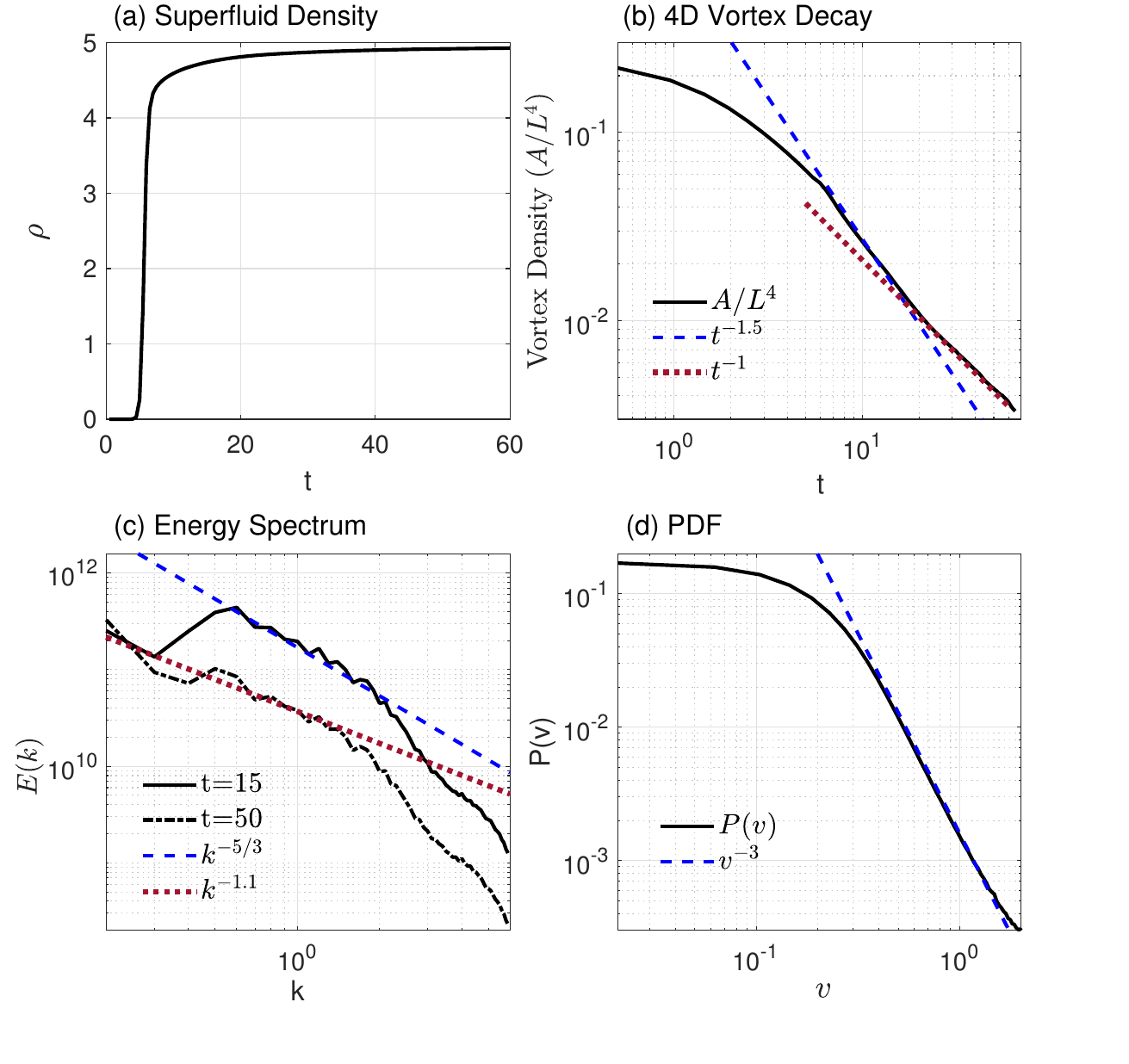}
\caption{(a) Superfluid density evolution after sudden quench.
(b) Decay of the vortex surface density showing two distinct regions: an initial quasi-classical decay region scaling as $t^{-1.5}$, followed by a transition to ultra-quantum turbulence scaling as $t^{-1}$.  
(c) Corresponding energy spectra measured in the steady decay stage, exhibiting a transition from the Kolmogorov $k^{-5/3}$ spectrum at higher vortex densities to a $k^{-1.1}$ spectrum. 
(d) Probability distribution function (PDF) of velocity magnitudes during the steady decay stage, displaying a non-Gaussian power-law tail \( P(v) \sim v^{-3} \).
}
\label{Trubulencedata}
\end{figure}

In our quench simulations, we initialize the system with a vanishing condensate field $\phi = 0$ with small thermal fluctuations
$\langle \delta\phi(\mathbf{r}) \delta\phi^*(\mathbf{r}') \rangle = 10^{-6} \delta(\mathbf{r} - \mathbf{r}')$.    
The chemical potential is then linearly ramped across the critical point, and we record the number of topological defects at the freeze-out time for various quench rates. As shown in Fig.\ref{4DKZM}(a), for slow quenches (large $\tau_Q$), the results exhibit excellent agreement with the predicted $-1/2$ power-law scaling, confirming the validity of the Kibble-Zurek mechanism in higher dimensions. Moreover, this scaling behavior demonstrates the uniqueness of 2D vortex surfaces as the only relevant topological defects in 4D superfluids.

In the fast quench region(small $\tau_Q$), however, the conventional Kibble-Zurek mechanism breaks down. Due to the rapid change of parameters, the freezing time is no longer determined by quenching, but only reflects the final parameter. In this region, the number of topological defects is no longer governed by the equilibrium correlation length at the freeze-out point, but rather by the instantaneous correlation length associated with the final quench value. This leads to an alternative power-law scaling, sometimes referred to as the 'fast-quench KZM', where the defect density scales as \cite{Zeng_2023,Saito_2007,Yang_2023Universal,Supplementary_material}
\begin{equation} 
n \sim \frac{\xi(\mu_f)^{d}}{\xi(\mu_f)^{D}} \sim \mu_f^{(D-d)\nu}=\mu_f^{1} 
\end{equation}
As shown in Fig.\ref{4DKZM}(b), our numerical results agree well with this prediction.

After a sudden quench with $\mu_f=5$, the background density rapidly increases to near the equilibrium value $\rho=\mu/g$, leading to the formation of a highly tangled quantum vortex state that characterizes quantum turbulence. Following the onset of a steady decay region, we measure the decay rate of the vortex surface density. As shown in Fig.~\ref{Trubulencedata}, the decay exhibits two distinct regions depending on vortex density: at high density, the decay follows the quasi-classical turbulence with decay rate $-1.5$, whereas at lower density, it transitions to the ultra-quantum Vinen turbulence characterized by a decay exponent of $-1$ \cite{Vinen_2002,Vinen2001,Walmsley_2007,Smith_1993,Bradley_2006}.

As previously discussed, the decay of vortex surfaces in four dimensions proceeds through mechanisms such as surface splitting, shrinking, and possible surface wave. These processes are analogous to vortex reconnections, ring shrinking, and Kelvin wave excitations in 3D turbulence, respectively, making the emergence of the same decay exponent physically reasonable.

In addition, we observe a Kolmogorov-like energy spectrum in the high-density region, characterized by a power-law scaling of
\begin{equation}
E(k) \sim k^{-5/3},
\end{equation}
providing strong evidence for the existence of a direct energy cascade and self-similarity in four-dimensional quantum turbulence~\cite{Kobayashi_2005,Kobayashi_2006,Yepez_2009,Yang_2025quantumturbulence,Zeng_2025}. However, in the low-density Vinen turbulence region, the spectrum transitions to a weaker $k^{-1.1}$ scaling, signaling the onset of purely quantum dynamics.

Furthermore, the probability distribution function (PDF) of velocity magnitudes exhibits a non-Gaussian power-law tail with exponent
\begin{equation}
P(v) \sim v^{-3},
\end{equation}
which is a hallmark feature distinguishing quantum turbulence from classical turbulence \cite{DeWaele_1994,Paoletti_2008}.
Together, these observed power-law behaviors confirm the existence of quantum turbulence in four dimensions and demonstrate its universal properties.

{\it Conclusion.} We have successfully performed numerical simulations and dynamical analyses of four-dimensional quantum vortices and turbulence. Several key findings were revealed. First, quantum vortices in four dimensions behave as surface-like structures, and no isolated vortices are observed; instead, they form a fully connected vortex surface. This behavior was previously predicted \cite{McCanna_2024}, and our results provide the first numerical confirmation. The decay of vortices occurs through surface splitting, which serves as the higher-dimensional analogue of vortex reconnection in three dimensions, leading to a reduction in the topological number (genus).
Due to the highly complex geometry of the vortex surface at early times, it is difficult to quantitatively determine the genus during the initial stages of evolution. Therefore, a precise quantitative study of genus decay is currently limited. Investigating whether the genus of 4D vortex surfaces follows a universal power-law decay remains an important direction for future work.

Through linear quenches across the phase transition, we have verified the applicability of both slow and fast versions of the Kibble-Zurek mechanism in four-dimensional system. Furthermore, in the turbulent regime that follows, we have identified multiple universal scaling laws. These signatures collectively suggest the presence of Kelvin-wave-like excitations in 4D quantum turbulence, which transfer energy from large to small scales. Developing a Bogoliubov theoretical framework to identify and analyze such excitations and their dispersion relations would be a interesting direction for future theoretical investigations. 

Finally, from the experimental perspective, recent advances in higher-dimensional physics in cold atom systems provide a promising avenue for realization \cite{Bouhiron_2024,Ozawa_2019}. In particular, synthetic dimensions engineered via fast shaking of optical lattices or coupling of internal atomic states are expected to enable the observation of quantum vortices in synthetically constructed four-dimensional condensates \cite{Madani_2025}.

{\it Acknowledgements.}
We would like to thank Makoto Tsubota and Xin Wang for valuable discussions.
This work is supported by the JSPS KAKENHI via Grant Number JP24K00569.


%

\onecolumngrid
\appendix 
\clearpage
\section*{Appendix}
\section{Numerical Method for the 4D Gross–Pitaevskii Equation}

As described in the main text, we adopt the natural higher-dimensional extension of the Gross–Pitaevskii equation (GPE) to model a four-dimensional (4D) weakly interacting Bose gas at low temperature. To incorporate thermal fluctuations, we employ the stochastic Gross–Pitaevskii equation (SGPE), which takes the form:
\begin{equation}
(i - \gamma)\frac{\partial \phi({\bf r},t)}{\partial t} 
= \left(-\frac{1}{2}\nabla^2 -\mu + g|\phi({\bf r},t)|^2\right)\phi({\bf r},t)+ \eta({\bf r},t),
\end{equation}
where ${\bf r} = (x, y, z, w)$ is the four-dimensional spatial coordinate, $\phi$ is the condensate wave function, $\gamma$ is the phenomenological dissipation parameter, $\mu$ is the chemical potential, $g$ is the interaction strength, and $\eta({\bf r}, t)$ is the complex Gaussian noise term representing thermal fluctuations. The noise satisfies the fluctuation-dissipation relation:
\begin{equation}
\langle \eta({\bf r}, t)\eta^*({\bf r}', t') \rangle = 2\gamma T\, \delta({\bf r} - {\bf r}')\, \delta(t - t'),
\end{equation}
where $T$ is the effective temperature.

The condensate wave function can be expressed in the Madelung form:
\begin{equation}
\phi({\bf r},t) = \sqrt{\rho({\bf r},t)}\, e^{i\theta({\bf r},t)},
\end{equation}
where $\rho$ is the local density and $\theta$ is the phase. In four dimensions, vortex solutions still exist as topological defects, characterized by quantized phase winding:
\begin{equation}
\oint \nabla \theta \cdot d\mathbf{l} = 2\pi n\kappa,
\end{equation}
where $\kappa$ is the circulation quantum and $n \in \mathbb{Z}$ is the winding number.

We discretize a four-dimensional cubic domain of size $L^4 = 30^4$ using $100^4$ grid points. The time evolution is performed with a fixed time step of $h = 10^{-4}$.
To implement spatial derivatives efficiently and accurately, we impose periodic boundary conditions along all four spatial directions and employ a Fourier spectral method. This allows us to compute the Laplacian operator $\nabla^2$ in Fourier space with high precision, taking full advantage of the periodic geometry of the simulation domain.

As the initial condition, we begin with a uniform condensate wave function $\phi = \sqrt{\mu/g}$, and imprint vortices using a phase modulation method. Specifically, for each vortex, we modify the phase by a factor of the form:
\begin{equation}
\phi \rightarrow \phi \cdot \exp(i \theta_i)
\end{equation}
with
\begin{equation}
 \theta_i(a, b) = s_i \arctan\left(\frac{a - a_i}{a - b_i}\right),
\end{equation}
where $(a_i, b_i)$ specifies the coordinates at which the $i$-th vortex is located in a plane perpendicular to the $(a, b)$ plane, and $s_i = \pm 1$ denotes the circulation sign (positive or negative) of the vortex.

Because each vortex is imprinted in a plane perpendicular to the selected coordinate pair, the resulting vortex surface lies along the remaining two directions—i.e., the vortex sheet is parallel to the other two axes. We choose to imprint one positive and one negative vortex surface randomly in each of the six independent 2D planes in 4D space: $(xy)$, $(xz)$, $(xw)$, $(yz)$, $(yw)$, and $(zw)$.

The system is then allowed to evolve freely without external driving, using the standard fourth-order Runge–Kutta method for time integration.

\section{4D Vortex surface visualization}

As described in the main text and illustrated in Fig.1, the visualization method we use projects four-dimensional vortex surfaces into three-dimensional space, allowing the structural and topological features of the 4D configuration to be intuitively displayed. To clarify this method, we illustrate it using a projection from 3D vortex lines into 2D space.

As shown in Fig. \ref{3dProjection2d}, we first slice the 3D vortex line configuration at a fixed time. The four small panels at the top display cross-sectional slices in the 
$x-y$ plane at different $z$ values. These slices are then sequentially stock together along a chosen direction in the $x-y$ plane, with physical spacing $R_z/n_z$, to form the 2D projection shown in panel (e). Although this 2D image inevitably loses some spatial information compared to the full 3D rendering in panel (f), and different projection directions (i.e., stocking along different directions) lead to different geometric deformations, the underlying topological structure remains unchanged—similar to how a higher-dimensional object can cast various lower-dimensional shadows depending on the viewing angle.

Moreover, the 3D visualization in Fig.1 (a Projection of the 4D superfluid) partially compensates for the loss of spatial information through the use of transparency and color, which allow us to distinguish whether vortex structures overlap in space. Therefore, this stitching method provides a lower-dimensional representation that still retains most of the essential topological and structural information.

\renewcommand{\thefigure}{S\arabic{figure}}
\setcounter{figure}{0}
\begin{figure}[h]
\includegraphics[width=14cm,trim=0 0 0 0]{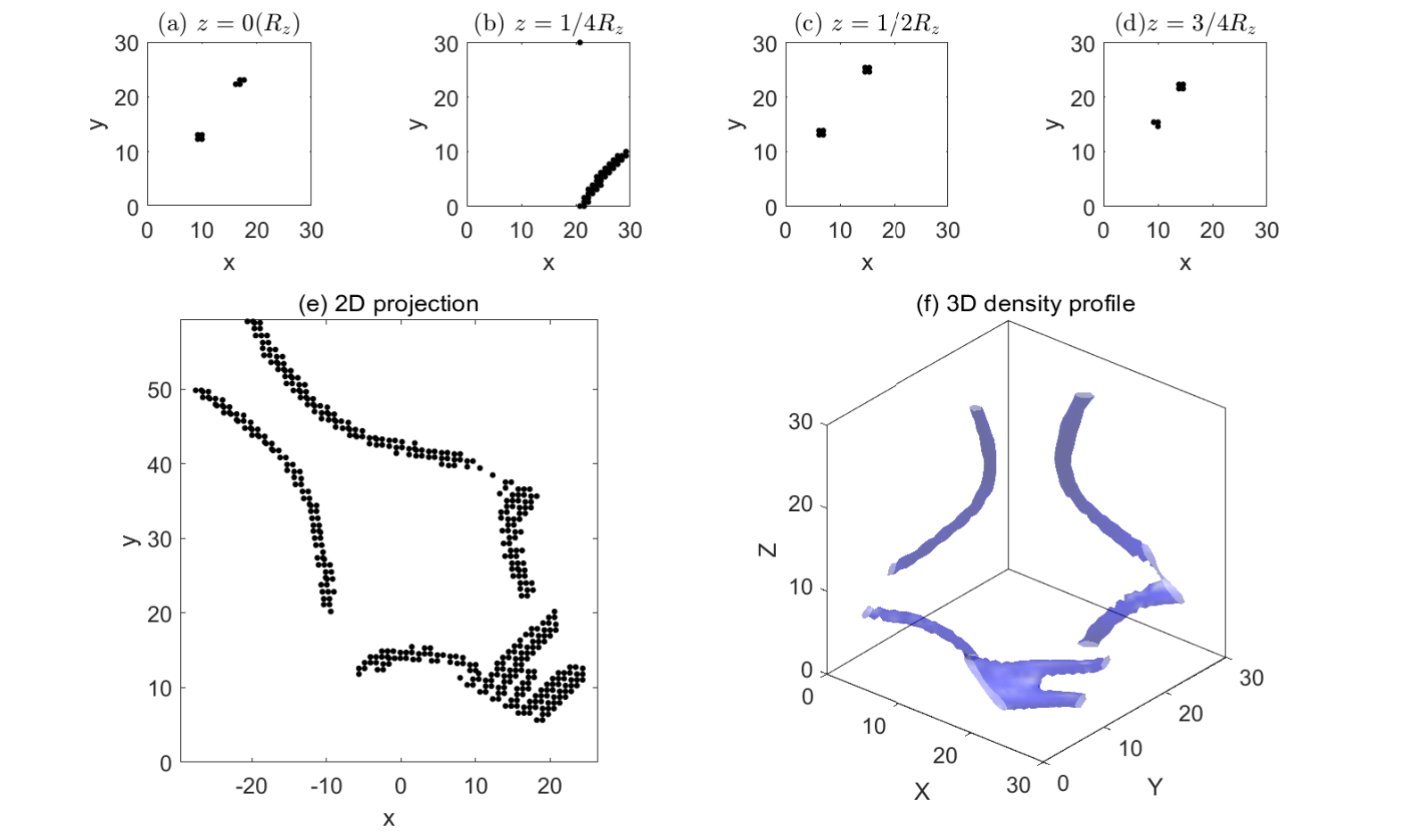}
\caption{Cross-sectional and reconstructed visualization of vortex structures in a three-dimensional quantum fluid. \\
Panels (a)--(d) show $xy$-plane slices of the vortex configuration at four different heights along the $z$-axis: $z = 0(R_z)$, $z = \frac{1}{4}R_z$, $z = \frac{1}{2}R_z$, and $z = \frac{3}{4}R_z$, respectively.\\
Panel (e) presents a 2D projection image constructed by stacking vortex points from all $z$ slices along the $x$ direction, illustrating the global topological pattern of the vortex network.\\
Panel (f) displays the original 3D vortex profile.
}
\label{3dProjection2d}
\end{figure}

\section{Kibble-Zurek Mechanism}

The Kibble-Zurek Mechanism (KZM) provides a universal framework for describing the non-equilibrium dynamics of systems undergoing continuous phase transitions at finite rates. Originally proposed in cosmology, it has been widely applied to condensed matter and quantum systems, including Bose-Einstein condensates.

Near the critical point $\lambda_c$, the system is characterized by a control parameter $\epsilon(t) = \lambda(t)-\lambda_c$, then the equilibrium correlation length $\xi$ and relaxation time $\tau$ diverge as:
\begin{equation}
\xi = \frac{\xi_0}{|\epsilon|^\nu}, \qquad \tau = \frac{\tau_0}{|\epsilon|^{z\nu}},
\end{equation}
where $\nu$ and $z$ are the correlation-length and dynamical critical exponents, respectively.

For a linear quench $\epsilon(t) =(\lambda_f-\lambda_c) t/\tau_Q+\lambda_c$, the system falls out of equilibrium near the critical point due to critical slowing down. 
The time for which the distance from the transition equals the relaxation time is called freeze-out time $\hat{t}$, the system is considered frozen for $|t| < \hat{t}$ and adiabatic elsewhere.

The freeze-out time $\hat{t}$ can be obtained by Zurek’s simple argument $\tau(\hat{t}) = \hat{t}$, leading to:
\begin{equation}
\hat{t} \sim \tau_0 \left( \frac{\tau_Q}{\tau_0} \right)^{\frac{z\nu}{1 + z\nu}}, \qquad
\hat{\xi} \sim \xi_0 \left( \frac{\tau_Q}{\tau_0} \right)^{\frac{\nu}{1 + z\nu}}.
\end{equation}

\renewcommand{\thefigure}{S2}
\setcounter{figure}{0}
\begin{figure}[h]
\includegraphics[width=13cm,trim=0 70 0 30]{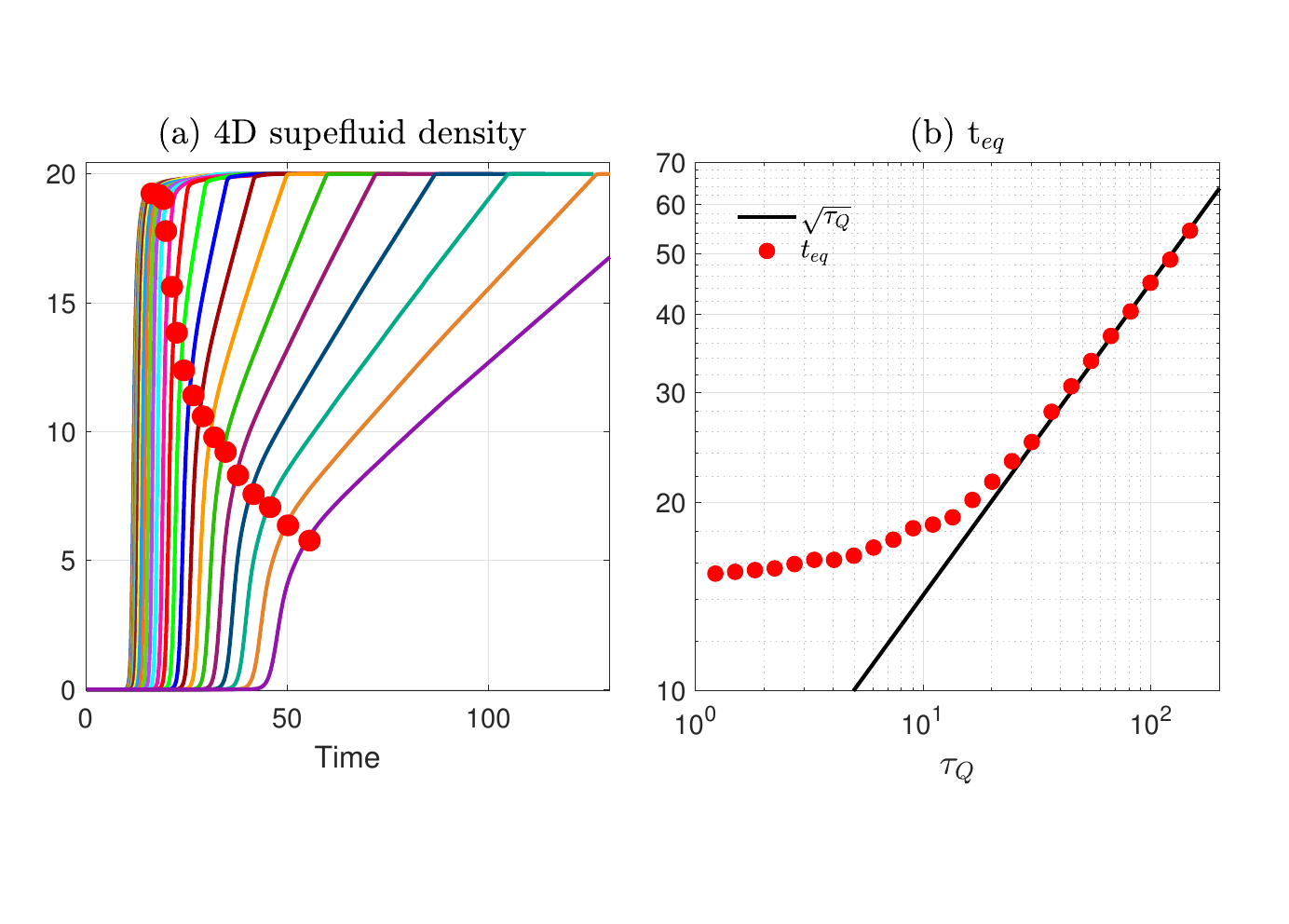}
\caption{Kibble-Zurek scaling of the equilibration time in four-dimensional condensate growth dynamics.  
(a) Time evolution of the condensate norm for various quench times $\tau_Q = \exp(0.2 \times j)$ with $j = 1, 2, \dots, 25$. Each colored curve corresponds to a different $\tau_Q$, showing a transition from early-time exponential growth to late-time linear behavior. The red dots mark the equilibration time $t_{\mathrm{eq}}$ for each quench.  
(b) Scaling of the extracted equilibration times $t_{\mathrm{eq}}$ (red circles) as a function of quench time $\tau_Q$ in a log-log plot. The solid black line indicates the Kibble-Zurek prediction $t_{\mathrm{eq}} \propto \sqrt{\tau_Q}$, which is in good agreement with the numerical data in the adiabatic regime.
}
\label{teq}
\end{figure}

Figure~\ref{teq} demonstrates the consistency of our results with the expected scaling, as the evolution curves for different quench times $\tau_Q$ collapse under proper rescaling. The numerical data reveal that the condensate norm undergoes an initial exponential rise followed by a linear-in-time growth phase. We designate the crossover between these two regions as the equilibration time $t_{\mathrm{eq}}$ (highlighted by red points), which characterizes the onset of post-critical dynamics. The dependence $t_{\mathrm{eq}} \propto \sqrt{\tau_Q}$, anticipated by the Kibble-Zurek framework, is clearly confirmed by the log-log scaling behavior shown in the right panel in the slow quench (large $\tau_Q$) region.

As a result, the average $d$-dimensional defect density in a $D$-dimensional system scales as:
\begin{equation}
n \sim \frac{\hat{\xi}^d}{\hat{\xi}^D} \sim \tau_Q^{-\frac{(D-d)\nu}{(1 + z\nu)}}.
\end{equation}

This framework provides the theoretical basis for the spontaneous formation of vortex structures in our simulations following a sudden quench of the chemical potential,  which is quantitatively verified in our four-dimensional setting.

However,  in the limit of rapid quenches, the system is governed by a single freeze-out time
\begin{equation}
\hat{t} \sim \tau(\lambda_f) \propto \epsilon_f^{-z\nu},
\end{equation}
which becomes independent of the quench time $\tau_Q$. Here, $\epsilon_f=\lambda_f-\lambda_c$ characterizes the distance from criticality at the final value of the control parameter.

As a result, the domain size is determined by the equilibrium correlation length at freeze-out,
\(\hat{\xi} = \xi(\lambda_f)\), leading to a plateau in the defect density for sufficiently fast quenches. This allows us to predict the scaling of the saturated defect density with the final control parameter:
\begin{equation}
n \sim \frac{\xi(\lambda_f)^d}{\xi(\lambda_f)^D} \propto \epsilon_f^{(D-d)\nu}.
\end{equation}

\section{Vortex Surface Density Calculation}

The concept of vortex density plays a central role in quantifying the degree of turbulence in quantum fluids. In three-dimensional systems, vortex lines appear as phase singularities in the complex order parameter $\phi(\mathbf{r}) = \sqrt{\rho(\mathbf{r})}e^{i\theta(\mathbf{r})}$, and the vortex line density is defined as the total length of vortex lines per unit volume. Numerically, this length is computed by detecting plaquettes where the phase $\theta$ winds by $2\pi$, and integrating over vortex segments oriented by the local vorticity vector $\boldsymbol{\omega} = \nabla \times \nabla\theta$.

In four-dimensional space, however, quantum vortices manifest as two-dimensional topological defects---surface-like structures rather than line singularities. Consequently, the vortex density must be generalized from a line length per unit volume to a surface area per unit 4D volume. Unlike the three-dimensional case, there is no natural vorticity vector in four dimensions. Instead, vorticity is described by a rank-2 antisymmetric tensor field:
\begin{equation}
\Omega_{\mu\nu} = \partial_\mu v_\nu - \partial_\nu v_\mu, \qquad \mu,\nu \in \{x, y, z, w\},
\end{equation}
which can be interpreted as a differential 2-form. There are six independent components, each corresponding to a specific 2D coordinate plane.

To compute the vortex surface density, we begin by identifying vortex core points from the complex field $\phi(\mathbf{r})$, using standard $2\pi$ phase winding detection on each of the six 2D planes: $xy$, $xz$, $xw$, $yz$, $yw$, and $zw$. The union of all detected points forms a discrete sampling of the underlying vortex surfaces in the 4D grid.

At each identified vortex point, we estimate the local area of the vortex surface element using a geometric method. Specifically, we find two nearest neighboring vortex points in non-collinear directions to construct local tangent vectors $\mathbf{t}_1$ and $\mathbf{t}_2$. These define the local surface via a wedge product:
\begin{equation}
\mathcal{A} = \mathbf{t}_1 \wedge \mathbf{t}_2,
\end{equation}
which is a bivector representing the oriented area element. The corresponding surface area is given by the norm of the bivector:
\begin{equation}
A_{\text{segment}} = \|\mathbf{t}_1 \wedge \mathbf{t}_2\| = \sqrt{|\mathbf{t}_1|^2 |\mathbf{t}_2|^2 - (\mathbf{t}_1 \cdot \mathbf{t}_2)^2},
\end{equation}
which is the area of the parallelogram spanned by $\mathbf{t}_1$ and $\mathbf{t}_2$ in four-dimensional space.

Summing over all such local segments yields the total vortex surface area:
\begin{equation}
A_{\mathrm{total}} = \sum_i A_{\mathrm{segment}}^{(i)}.
\end{equation}

Finally, the vortex surface density is defined by normalizing the total area by the full four-dimensional system volume $V = L^4$:
\begin{equation}
n_{\mathrm{vortex}} = \frac{A_{\mathrm{total}}}{L^4}.
\end{equation}

This bivector-based method captures the geometric essence of vortex surfaces in four-dimensional quantum turbulence and offers a physically consistent generalization of vortex line density to higher dimensions.

\section{Energy Spectrum}

To characterize the turbulent energy distribution across spatial scales, we compute the kinetic energy spectrum from the superfluid velocity field.

The velocity field is defined as:
\begin{equation}
\mathbf{u}({\bf r}) = \frac{i}{2} \frac{\phi^* \nabla \phi - \phi \nabla \phi^*}{|\phi|^2},
\end{equation}
where $\phi({\bf r})$ is the complex condensate wave function and $\nabla$ denotes the gradient operator in four-dimensional space.

The velocity field is first Fourier-transformed:
\begin{equation}
\mathbf{u}({\bf k}) = \mathcal{F}[\mathbf{u}({\bf r})],
\end{equation}
where $\mathbf{k} = (k_x, k_y, k_z, k_w)$ is the four-wavevector in Fourier space.

The kinetic energy spectrum $E_k(k)$ is computed by integrating the squared velocity amplitude over a thin shell in $k$-space with magnitude between $k$ and $k + \Delta k$:
\begin{equation}
E_k(k) = \frac{1}{2} \int_{k \leq  |\mathbf{k}'| < k + \Delta k} \rho \left| \mathbf{u}(\mathbf{k}') \right|^2 \, d^4k'.
\end{equation}

Here, $|\mathbf{k}'| = \sqrt{k_x^2 + k_y^2 + k_z^2 + k_w^2}$ is the magnitude of the 4D wavevector, and the integral is performed over all modes lying within a 4D spherical shell in Fourier space. The resulting spectrum $E_k(k)$ captures the distribution of kinetic energy as a function of length scale and reveals scaling laws associated with quantum turbulence.

\end{document}